\documentclass[12pt]{article}
\usepackage{amssymb,latexsym,amsmath}
\usepackage{amsfonts, graphics}
\newcommand{\R}{\mathbb R}

\def\open#1{\setbox0=\hbox{$#1$}
\baselineskip = 0pt
\vbox{\hbox{\hspace*{0.4 \wd0}\tiny $\circ$}\hbox{$#1$}}
\baselineskip = 11pt\!}
\def\fn{\open{f}}

\def\be{\begin{equation}}
\def\ee{\end{equation}}
\def\bea{\begin{eqnarray}}
\def\eea{\end{eqnarray}}
\def\beas{\begin{eqnarray*}}
\def\eeas{\end{eqnarray*}}

\def\supp{\mathrm{supp}\,}

\begin{document}

\title{Global existence for the spherically symmetric Einstein-Vlasov system
with outgoing matter}

\author{H{\aa}kan Andr\'{e}asson\\
        Department of Mathematics, Chalmers,\\
        S-41296 G\"oteborg, Sweden\\
        email: hand@math.chalmers.se\\
        \ \\
        Markus Kunze\\
        Fachbereich Mathematik\\
        Universit\"at Duisburg-Essen\\
        D-45117 Essen, Germany\\
        email: markus.kunze@uni-due.de\\
        \ \\
        Gerhard Rein\\
        Fakult\"at f\"ur Mathematik und Physik\\
        Universit\"at Bayreuth,\\
        D-95440 Bayreuth, Germany\\
        email: gerhard.rein@uni-bayreuth.de}

\maketitle

\begin{abstract}
We prove a new global existence result for the asymptotically
flat, spherically symmetric Einstein-Vlasov system
which describes in the framework of general relativity
an ensemble of particles which interact by gravity.
The data are such that initially all the particles 
are moving radially outward and that this property
can be bootstrapped. The resulting non-vacuum spacetime
is future geodesically complete.
\end{abstract}
\section{Introduction}
\setcounter{equation}{0}
Global properties of non-vacuum spacetimes are one of the central
themes of the mathematical analysis of general relativity.
A matter model which has proven particularly suitable for
such an analysis is the collisionless gas as described by the Vlasov
equation. Here the matter content of spacetime is represented
by a large ensemble of particles
with a density function $f$ on phase space, i.e., on the
mass shell of the tangent bundle of the spacetime manifold.
The particles move along time-like geodesics which is equivalent
to $f$ satisfying the corresponding continuity equation, i.e., 
the Vlasov equation, and the energy momentum tensor is then expressed 
in terms of $f.$ 

For a general introduction to the resulting Einstein-Vlasov
system we refer to \cite{An1} and \cite{Rl7}. A major result
for the asymptotically flat case of this system is a global 
existence result for small, spherically symmetric initial data
\cite{RR1}. 
In addition, there is numerical
evidence \cite{AR1,OC,RRS2} that at least in the spherically symmetric case
the weak cosmic censorship conjecture holds: The gravitational collapse
of regular initial data leads to a black hole rather than 
a naked singularity. 

In view of the small data result mentioned above and the lack of proof of global 
existence
for general data (in Schwarzschild coordinates or in maximal-areal coordinates) 
it is natural to ask if one can find an open set of initial
data which are not small in the sense of \cite{RR1} but for which global existence
holds. In the present paper we do find initial data with this property by
establishing global existence for data which are such that initially all 
the particles in the ensemble move radially outward
sufficiently fast so that they can be shown to keep on moving outward
for all future time. This configuration results in estimates
on the geometry of spacetime which imply that it is future geodesically 
complete. 

The basic mechanism of the proof is completely different
from the one for small initial data which relied on the
dispersive properties of the Vlasov matter model when the
fields are small. Since its actual implementation for the 
Einstein-Vlasov system quite hides this mechanism, it should
be useful to explain it first in the case of the much simpler
Vlasov-Poisson system which is the Newtonian limit of the
Einstein-Vlasov system. Due to the spherical symmetry the
maximal force experienced by a particle at distance $r$
from the origin is $-M/r^2$ in the Vlasov-Poisson case,
where $M>0$ is the total mass of the ensemble. Hence
along any particle trajectory
\[
\frac{d}{dt} \left(\frac{1}{2} w^2 - \frac{M}{r}\right) =
 w \dot w + \frac{M}{r^2}\dot r = w\,\left(\dot w + \frac{M}{r^2}\right)
\geq  0,
\]
as long as its radial velocity $\dot r = w= x\cdot v/r \geq 0$.
Hence
\[
\frac{1}{2} w^2(t) - \frac{M}{r(t)} \geq \frac{1}{2} w^2(0) - \frac{M}{r(0)}
\]
and
\[
\frac{1}{2} w^2(t) \geq \frac{1}{2} w^2(0) - \frac{M}{r(0)}
\]
on any time interval on which $w(t)$ remains non-negative. 
Now let $w_->0$ be an initial lower bound for the radial velocities
of the particles in the ensemble, $r_->0$ an initial lower bound for 
their distance from the origin,
and assume that 
\[
W_- := \frac{1}{2} w_-^2 - \frac{M}{r_-} > 0.
\]
Then as long as a particle is moving outward,
\[
w(t) > W_-,\ r(t) > r_- + W_- t.
\]
But this implies that all the particles keep moving outward for
all future time. For the Vlasov-Poisson system 
global existence and uniqueness of smooth solutions
has been established for general initial data 
\cite{Pf,LP}, but the above argument can easily be extended
to yield global existence for spherically symmetric initial data as 
specified above.

For the spherically symmetric Einstein-Vlasov system we find
a substitute for the pseudo-energy 
$\frac{1}{2} w^2 - \frac{M}{r}$ used above, but when taking its
derivative along a particle trajectory it turns out to be
important to use suitable coordinates on spacetime,
namely maximal-areal coordinates, in order
to see that under suitable circumstances this quantity
does not decrease along radially outgoing trajectories.
Once this is shown a bootstrap argument implies
that all the particles keep moving radially outward
on the future directed maximal interval of existence
of the solution. We can then use a result from
\cite{Rl7} to conclude that the solution is
global in the time coordinate used, and the obtained control of
the metric quantities turns out to be sufficient
to establish future geodesic completeness.
\section{The Einstein-Vlasov system}
\setcounter{equation}{0}
We consider the system in the spherically
symmetric, asymptotically flat case and write the metric
in the following form:
\[
ds^{2}=-(\alpha^2-a^2\beta^2)dt^2+2a^2\beta dtdr+a^2 dr^2
+ r^2\left(d\theta^2 + \sin^2\theta\, d\phi^2\right).
\]
Here the metric coefficients $\alpha, \beta$, and $a$ depend on
$t\in \R$ and $r\geq 0$,
$\alpha$ and $a$ are positive, and the polar angles $\theta\in[0,\pi]$
and $\phi\in[0,2\pi]$ parameterize the unit sphere.
The radial coordinate $r$ is thus the area radius.
Let $K^{a}_{b}$ be the second fundamental form and define
\be \label{kdef}
   \kappa=K^{\theta}_{\theta}=\frac{\beta}{r\alpha}.
\ee
By imposing the maximal gauge condition, which means that
each hypersurface of
constant $t$ has vanishing mean curvature, we obtain
the following field equations where subscripts denote partial derivatives
with respect to $r$ or $t$:
\begin{eqnarray}
   a_{r} & = & \frac{3}{2}a^{3}r \kappa^2+4\pi
   r a^{3}\rho+\frac{a}{2r}(1-a^{2}), \label{ee1}\\
   \kappa_{r} & = & -\frac{3}{r}\kappa-4\pi a\jmath, \label{ee2}\\
   a_{t} & = & 2\alpha a \kappa+(a\beta)_{r}, \label{ee3}\\
   \displaystyle\alpha_{rr} & = & \alpha_{r}\left(\frac{a_{r}}{a}-\frac{2}{r}\right)
   +\frac{2\alpha}{r^2}\left(2r\frac{a_{r}}{a}+a^2-1\right)+4\pi
   a^2\alpha(S-3\rho). \label{ee4}
\end{eqnarray}
The Vlasov equation takes the form
\begin{equation} \label{vlasov}
\partial_{t}f+\left(\frac{\alpha w}{a E}-\beta\right)\partial_{r}f
   +\left(-\frac{\alpha_{r}  E}{a}-2\alpha \kappa w+ \frac{\alpha L}
   {a r^3  E}\right)\partial_{w}f=0, 
\end{equation}
where
\begin{equation}\label{E}
   E=E(r,w,L)=\sqrt{1+w^{2}+L/r^{2}}.
\end{equation}
The variables $w$ and $L$ can be thought of as the momentum in the
radial direction and the square of the angular momentum
respectively, see \cite{R} for more details. The matter
quantities which appear in the field equations are defined by
\begin{eqnarray}
   \rho(t,r) & = & \frac{\pi}{r^{2}}
   \int_{-\infty}^{\infty}\int_{0}^{\infty} E f(t,r,w,L)
   \,dL\,dw, \label{rho-def}
   \\ \jmath(t,r) & = & \frac{\pi}{r^{2}}\int_{-\infty}^{\infty}\int_{0}^{\infty}
   w f(t,r,w,L)\,dL\,dw, \label{j-def}
   \\ S(t,r) & = & \frac{\pi}{r^{2}}\int_{-\infty}^{\infty}\int_{0}^{\infty}
   \frac{w^2 + L /r^{2}}{E}\, f(t,r,w,L)\,dL\,dw. \label{S-def}
\end{eqnarray}
We impose the following boundary conditions which ensure asymptotic flatness
and a regular center:
\begin{equation}\label{bdryc}
   a(t,0)=a(t,\infty)=\alpha(t,\infty)=1.
\end{equation}
The equations (\ref{ee1})--(\ref{S-def}) together with the relation (\ref{kdef})
and the boundary conditions
(\ref{bdryc}) constitute the Einstein-Vlasov system for a spherically
symmetric, asymptotically flat spacetime
in maximal-areal coordinates.

First we note that the phase space density $f$ is constant along
solutions of the characteristic system
\begin{eqnarray}
   \dot r & = & \frac{\alpha(\tau,r) w}{a(\tau,r) E}
   -\beta(\tau,r), \label{charsys1} \\
   \dot w & = & -\frac{\alpha_{r}(\tau,r) E}{a(\tau,r)}
   -2\alpha(\tau,r) \kappa(\tau,r) w
   +\frac{\alpha(\tau,r) L}{a(\tau,r) r^3 E}, \label{charsys2} \\
   \dot L & = & 0, \label{charsys3}
\end{eqnarray}
of the Vlasov equation. If $\tau\mapsto (R,W,L)(\tau,t,r,w,L)$
denotes the solution of the characteristic system with
$(R,W,L)(t,t,r,w,L)=(r,w,L)$ then
\[ f(t,r,w,L)=\fn((R,W,L)(0,t,r,w,L)), \]
with $\fn = f_{|t=0}$ the initial data for $f$.
In particular, $f\geq 0$ provided this is true for the initial data
as we will assume throughout.
Next we notice that (\ref{ee2}) can be rewritten as
\[
   \left(r^3 \kappa\right)_r = - 4 \pi r^3 a \jmath,
\]
and upon integration
\begin{equation}\label{kap-form}
   \kappa(t, r)=-\frac{4\pi}{r^3} \int_0^r a(t, s)\jmath(t, s) s^3\,ds.
\end{equation}
The Hawking mass $m$ is given by
\begin{equation}\label{m}
   m=\frac{r}{2}\left(1+\frac{\beta^2}{\alpha^2}-\frac{1}{a^2}\right).
\end{equation}
We also introduce the quantity
\begin{equation}\label{mudef}
   \mu=\frac{r}{2}\left(1-\frac{1}{a^2}\right)
\end{equation}
and note that by (\ref{ee1}), $\mu$ can be expressed as
\begin{equation}\label{murewr}
   \mu(t,r)=\int_{0}^{r}\left(4\pi \rho(t,s)
   +\frac{3}{2}\kappa^2(t,s)\right) s^2\,ds.
\end{equation}
Assuming that the matter is compactly supported
initially and hence also for later times, (\ref{kap-form}) implies that
$\kappa(t,r)\sim r^{-3}$ for $r$ large. Hence the limits
as $r$ tends to $\infty$ of $m$ and $\mu$ are equal,
so that the ADM mass $M$ is
\begin{equation}\label{M-def}
   M=\int_0^{\infty}\left(4\pi \rho(t,r)+\frac{3}{2}
   \kappa^2(t,r)\right) r^2 dr.
\end{equation}
The ADM mass is conserved, and $\mu(t, r)\le M$.

Finally we notice that using (\ref{ee1}) the second order equation 
(\ref{ee4}) can be rewritten in the form
\[
   \left(\frac{r^2}{a}\alpha_r \right)_r
   =4\pi r^2 a \alpha (\rho + S) + 6 r^2 a \alpha \kappa^2,
\]
and upon integration,
\begin{equation}\label{alphar1}
   \alpha_r(t, r)=\frac{a(t, r)}{r^2}\int_{0}^{r}\left(4\pi\alpha a (\rho+S)+6 a \alpha
   \kappa^2\right) s^2\,ds.
\end{equation}
Thus $\alpha$ is monotonically increasing outwards, and from (\ref{bdryc})
it follows that
\begin{equation}\label{alphaleq1}
0<\alpha\leq 1.
\end{equation}
A direct computation using 
(\ref{ee4}), (\ref{ee1}), (\ref{mudef}), and (\ref{murewr})
implies that
\[ 
\left(\frac{r^2}{a^2}\alpha_r - \alpha \mu \right)_r =
   4\pi r^2 S \alpha + \frac{9}{2} r^2 \kappa^2 \alpha
  -4\pi r^3\rho\alpha_r - \frac{3}{2} r^3 \kappa^2 \alpha_r. 
\]
With the boundary conditions
$(\alpha \mu)_{|r=0} = 0 = \left(r^2 a^{-2}\alpha_r\right)_{|r=0}$
this yields the identity
\begin{equation}\label{alphar2}
   \frac{r^2}{a^2(t,r)}\,\alpha_r(t,r)
   =(\alpha \mu)(t,r)
   +\int_0^r\left[4 \pi S \alpha + \frac{9}{2} \kappa^2 \alpha
   -4\pi s \rho \alpha_r - \frac{3}{2} s \kappa^2 \alpha_r\right] s^2 ds,
\end{equation}
which turns out to be very convenient in what follows.
\section{The main result}
\setcounter{equation}{0}
Let $\fn\,$ be some initial data of ADM mass $M$ such that
\[ 
\supp \fn \subset ]r_-, r_+[\times [w_-, \infty[\times [0, L_+]\,. 
\]
We wish to specify these parameters in such a way that 
for the corresponding solution of the Einstein-Vlasov system
all particles have radial velocity $w > W_-$ with an additional 
parameter $W_->0$. To this end
we abbreviate
\begin{equation} \label{xieta-def}
\xi = \frac{2M}{r_-},\ \eta = \frac{W_-}{\sqrt{1+W_-^2+L_+/r_-^2}},
\end{equation}
and we require
that the above parameters satisfy the following conditions:
\begin{eqnarray} \label{master1}
&&
\left(2\min\left\{\frac{r_-}{r_+},\frac{1-3\xi}{\sqrt{1-\xi}+ \xi/2} \eta\right\} 
-1 -\frac{9}{4}\,\frac{\xi}{1-\xi}
  -\frac{3}{2}\,\frac{ \xi}{\sqrt{1-\xi}} 
-\frac{L_+}{W_- r_-^2}\right) W_-^2 \nonumber \\ 
&&   
\qquad\qquad -\left(1+\frac{L_+ }{W_- r_-^2} + \frac{9}{4}\,\frac{\xi}{1-\xi}\right)
\left(1+\frac{L_+ }{r_-^2}\right) -\frac{L_+ }{r_-^2}\frac{1}{1-\xi} > 0
\end{eqnarray}
and
\begin{equation} \label{master2}
w_-^2> \frac{1}{1-\xi}\left(W_-^2+\xi\left(1+\frac{L_+}{r_-^2}\right)\right).
\end{equation}
After stating our main result we will show that there exist parameters
satisfying these conditions, and that there exist initial data with
such parameters. 

\smallskip

\noindent
{\bf Theorem} {\em
Suppose that $0<r_-<r_+$, $w_->0$, $L_+>0$, $W_->0$, and $M>0$
satisfy the conditions (\ref{master1}) and (\ref{master2}),
and consider a solution $f$ on a maximal existence interval
$[0, T[$ such that the initial data $\fn=f_{|t=0}$ is compactly supported,
smooth, non-negative,
has ADM mass $M$, and satisfies
\begin{equation}\label{supp-as}
   \supp\fn\subset ]r_-, r_+[\times [w_-, \infty[\times [0, L_+]\,.
\end{equation}
Then $T=\infty$, and the resulting spacetime is future geodesically complete.
}

\smallskip

\noindent
The proof of this theorem is given in Section \ref{proof-sec} below.

To show that parameters $\xi, r_\pm, w_-, L_+, W_-$ exist which satisfy 
(\ref{master1}) and (\ref{master2}) we proceed as follows.
Let 
\[ 
D(\xi,\eta)=2 \frac{1-3\xi}{\sqrt{1-\xi}+ \xi/2} \eta -1 
  -\frac{9}{4}\,\frac{\xi}{1-\xi}
  -\frac{3}{2}\,\frac{ \xi}{\sqrt{1-\xi}},\quad\xi\in [0, 1[,\ \eta >0. 
\]
Since $D(0,1)=1$ and $D(1/3,1) < 0$ there exists some $\xi^\ast \in ]0,1/3[$ such that
$D(\xi,1)>0$ for $\xi\in [0, \xi^\ast[$ and $D(\xi^\ast,1)=0$; numerically, $\xi^\ast\approx 0.0994$.
So firstly we assume that
\[
   \xi<\xi^\ast.
\]
Secondly, we choose $r_+>r_->0$ such that
\[
   2 \frac{r_-}{r_+} > 1 + \frac{9}{4}\,\frac{\xi}{1-\xi}
   + \frac{3}{2}\,\frac{ \xi}{\sqrt{1-\xi}},
\] 
where we notice that the right hand side is increasing in $\xi$
and its value for $\xi=\xi^\ast$ is $\approx 1.405$ so that 
the inequality can be satisfied with $r_+>r_->0$.
Now we fix $L_+>0$ arbitrarily. Since $\eta \to 1$ as $W_- \to \infty$
we can choose $W_->0$ sufficiently large so that 
\[ 
D(\xi,\eta) - \frac{L_+ }{W_- r_-^2} > 0
\]
and
\[ 
   2 \frac{r_-}{r_+} > 1 + \frac{9}{4}\,\frac{\xi}{1-\xi}
   + \frac{3}{2}\,\frac{ \xi}{\sqrt{1-\xi}} + \frac{L_+ }{W_- r_-^2}.
\] 
Then the factor multiplying $W_-^2$ in the condition (\ref{master1})
is positive, and by further increasing $W_-$ if necessary
condition (\ref{master1}) is satisfied.
Finally, we take $w_->0$ sufficiently large so that
(\ref{master2}) holds as well.

Let us now fix a set of parameters $\xi, r_\pm, L_+, w_-, W_-$ such that
(\ref{master1}) and (\ref{master2}) are satisfied. We want to argue that
an initial condition with these parameter values does exist.
In order to do so we fix some compactly supported, smooth function $\open{g}\,$ such that
\[ 
\supp \open{g}\subset ]r_-, r_+[\times [w_-, \infty[\times [0, L_+]\,, 
\]
and we consider $\fn=A\open{g}\,$ for some amplitude $A>0$.
Quantities induced by $\fn$ will be denoted by the corresponding superscript,
and they depend on $A$. By continuity, we can make sure that
$\open{a}(r)\le 2$ for all $A>0$ sufficiently small;
the dependence of the initial data for the metric quantities on
the initial matter terms and hence on $\fn$ will be studied
in \cite{AKR}. Now let
\[ 
I=\int_0^\infty4\pi\open{\rho}(s) s^2\,ds. 
\]
Then by (\ref{kap-form}),
\[
|\open{\kappa}(r)|
\le\frac{4\pi}{r^3}\int_0^r\open{a}(s)\,|\open{\jmath}(s)|\,s^3\,ds
\le
\frac{8\pi}{r^2}\int_0^r\open{\rho}(s) s^2\,ds =\frac{2I}{r^2}\, .
\]
Since $\open{\kappa}(r)=0$ for $r\le r_-$
this implies that
\[
   I \leq M =  \int_{r_-}^\infty \left(4\pi\open{\rho}(r)
   +\frac{3}{2}\open{\kappa}^{\,\,2}(r)\right) r^2 dr
   \le I + \frac{6 I^2}{ r_-}.
\]
Since the quantity $I$ depends linearly on the amplitude $A$, $2M(\fn\,)/r_- =
\xi < \xi^\ast$ for $A>0$ sufficiently small, and arguing again by continuity 
any value of $\xi \in ]0,\xi^\ast[$ should be realized by adjusting $A$ properly.
Since $\fn$ has the same support as $\open{g\,}$, all conditions required
in the theorem then hold for $\fn$.
\section{Proof of the theorem}\label{proof-sec}
\setcounter{equation}{0}
{\em Step 1: The bootstrap argument.}\\
Our first and crucial step in the proof of the theorem is a
bootstrap argument which shows that the condition $w>W_-$
which holds on the support of $f$ initially will persist
on the maximal forward existence interval of the solution.

To begin with we show that
for all $t\in [0,T[$ and $r\geq r_-$,
\begin{equation} \label{gote}
a(t,r) < a_+=\frac{1}{\sqrt{1-\xi}}\ \mbox{and}\ 
\alpha(t,r) > \alpha_-= \frac{1-3\xi}{1-\xi};
\end{equation}
eventually we shall see that these bounds hold for all $r\geq 0$.
We recall that $\xi=2M/r_-$.
By (\ref{mudef}), 
\[
   \frac{1}{a^2(t, r)}=1-\frac{2\mu(t, r)}{r} > 1-\frac{2M}{r_-}=1-\xi
\]
for $r\geq r_-$;
the inequality is strict because the solution is non-trivial and hence
$\kappa$ does not vanish identically. 
By (\ref{alphar1}), (\ref{alphaleq1}),  and (\ref{M-def}),
\begin{equation} \label{alpha-r}
\alpha_{r}(t, r)
   \le\frac{a_+^2}{r^2}\int_{0}^{r}\left(4\pi (\rho+S)+6
   \kappa^2\right) s^2 ds < \frac{4 a_+^2 M}{r^2}\,. 
\end{equation}
Since $\alpha(t, \infty)=1$, this implies that
\[ 
\alpha(t, r) > 1-\frac{4 a_+^2 M}{r_-}
   =\frac{1-3\xi}{1-\xi}. 
\]
Next we define for $t\in [0, \infty[$,
\begin{eqnarray}
   R_-(t) & = &  r_- + \frac{\alpha_-}{a_+}
   \,\frac{W_-}{\sqrt{1+W_-^2 +L_+/r_-^2}}\,t,
   \label{R-def}
   \\ R_+(t) & = &   r_+ + \left(1 + \frac{a_+  M}{r_-}\right) t,
   \label{R+def}
\end{eqnarray}
as well as
\begin{eqnarray}\label{gamma-def}
   \gamma & = & \min_{t\in [0, \infty[}\frac{R_-(t)}{R_+(t)}
   =\min\left\{\frac{r_-}{  r_+},\frac{\alpha_-}{a_+(1+ a_+  M/r_-)}
   \frac{W_-}{\sqrt{1+W_-^2 +L_+/r_-^2}}\right\} \nonumber
   \\ & = & \min\left\{\frac{r_-}{  r_+},
   \frac{1-3\xi}{ \sqrt{1-\xi}+ \xi/2}
   \,\eta\right\},
\end{eqnarray}
where we recall the abbreviations in (\ref{xieta-def}) and (\ref{gote}).
Let $[0, t_0[$ denote the maximal time interval such that
for $t\in [0, t_0[$ and $(r, w, L)\in \supp f(t)$,
\begin{equation}\label{rwbd}
   R_-(t)<r<R_+(t),\quad w>W_-\,.
\end{equation}
By (\ref{master2}) and (\ref{supp-as}) this holds for $t=0$, and 
by continuity there is a maximal time interval $[0, t_0[$ with $t_0>0$
on which (\ref{rwbd}) persists.
Suppose that $t_0\in ]0, T[$. Then we must have equality at $t=t_0$
in at least one of the inequalities in (\ref{rwbd}). But in the following part
of the proof we show that (\ref{rwbd})
persists on $[0, t_0]$. Thus necessarily $t_0=T$.

Consider a characteristic such that $f(t, r(t), w(t), L) > 0$.
Denoting the parameter along the curve by $t$ it is calculated that
\bea\label{dds1}
   \frac{d}{dt}\left[\left(1-\frac{2 M}{r}\right) E^2\right]
   & = & - 2 \left(1-\frac{2 M}{r}\right)\frac{\alpha_r}{a} E w - 4 \frac{\beta}{r} w^2
   + 6 \frac{M}{r^2} \beta w^2 \nonumber \\
   & & {} + 2 \frac{L}{r^3} \beta
   - 6 \frac{M}{r^4} \beta L + 2 \frac{M}{r^2}\frac{\alpha}{a} E w
   - 2\frac{M}{r^2} \beta,
\eea
and we want to show that this quantity remains positive on the
time interval $[0,t_0[$.
By (\ref{j-def}) and (\ref{rwbd}), $\jmath\ge 0$ for $t\in [0,t_0[$. 
Hence (\ref{kdef}) and (\ref{kap-form})
lead to the estimate
\be\label{betaeq}
   \beta(t,r)=-\frac{\alpha(t,r)}{r^2}
   \,\int_0^r 4\pi\,a(t,s) \jmath(t,s) s^3 ds\leq 0,
\ee
and we can drop the fifth and seventh term in (\ref{dds1}). 
Using the estimate
\[ 
\frac{1}{a^2(t,r)} = 1-\frac{2\mu(t,r)}{r} \geq 1-\frac{2 M}{r} 
\]
and the fact that $\alpha_r\ge 0$, cf.\ (\ref{alphar1}), we find that
\be\label{dds2}
   \frac{d}{dt}\left[\left(1-\frac{2 M}{r}\right) E^2\right]
   \geq - 2 \frac{\alpha_r}{a^3} E w - 4 \frac{\beta}{r} w^2
   +\frac{2\beta}{r^2}\Big(3Mw^2+\frac{L}{r}\Big)
   + 2 \frac{M}{r^2}\frac{\alpha}{a} E w.
\ee
By (\ref{alphar2}) and in view of $\alpha(t, s)\le\alpha(t, r)$,
\[ \alpha_r(t,r)\leq\frac{M}{r^2} (\alpha a^2)(t,r)
   +\frac{(\alpha a^2)(t,r)}{r^2}
   \int_0^r\left(4 \pi S + \frac{9}{2} \kappa^2 \right) s^2 ds. \]
Inserting this into (\ref{dds2}) and using $a\geq 1$ and (\ref{betaeq}),
\bea\label{dds3}
   \frac{d}{dt}\left[\left(1-\frac{2 M}{r}\right) E^2\right]
   & \geq & \frac{2\alpha}{r^2}\biggl[\frac{2w^2}{r}\int_0^r 4 \pi \jmath s^3 ds
   -Ew\int_0^r\left(4 \pi S + \frac{9}{2} \kappa^2\right) s^2 ds\nonumber \\
   & & \qquad -\left(3 M w^2 +\frac{L}{r}\right) \frac{1}{r^2}
   \int_0^r 4 \pi a \jmath s^3\,ds\biggr];
\eea
notice that the first term coming from the above
estimate for $\alpha_r$ is exactly cancelled by the last term in (\ref{dds2}), 
which is crucial.
Now the estimates $w^2/E\le w$ and $E\ge 1$ imply that
\bea \label{Sest}
  S(t,r) & \leq & \jmath(t,r)+\frac{\pi}{r^{4}}\int_{-\infty}^{\infty}
  \int_0^{\infty}\frac{L}{E}\,f(t, w, r, L)\,dL\,dw
  \nonumber \\ & \le & \jmath(t, r)+\frac{\pi}{r^4}
  \,\frac{L_+}{W_-}\int_{-\infty}^{\infty}\int_0^{\infty}
  w\,f(t, w, r, L)\,dL\,dw \nonumber
  \\ & \leq & \left(1+\frac{L_+}{W_-}\frac{1}{r^2}\right)\,\jmath(t,r).
\eea
By (\ref{kap-form}), $\kappa(t,r)=0$ for $r\leq R_-(t)$, and
\begin{equation}\label{kappa-est}
   |\kappa(t,r)| \leq a_+ \frac{4\pi}{r^2} \int_0^r \jmath(t, s) s^2 ds
   \leq a_+ \frac{4\pi}{r^2} \int_0^r\rho(t, s) s^2 ds
   \leq \frac{a_+ M}{r^2}.
\end{equation}
Let
\[ 
J(t,r)=\int_0^r 4\pi \jmath(t,s) s^2 ds \geq 0. 
\]
Then using the fact that $J$ is increasing in $r$
and $R_-(t)\ge r_-$ for $t\in [0,t_0[$,
\begin{equation}\label{kintest}
  \int_0^r\kappa^2 s^2 ds
  \leq a_+^2 M\int_{R_-(t)}^r\frac{1}{s^2}
  \bigg(\int_0^s 4\pi \jmath  \tau^2 d\tau \bigg)\,ds
  \leq\frac{a_+^2 M}{R_-(t)}\,J
  \leq\frac{a_+^2  M}{r_-}\,J.
\end{equation}
By the definition of $\gamma$, see (\ref{gamma-def}),
\be\label{wrongjest}
   \frac{1}{r}\int_0^r 4\pi \jmath(t,s)\,s^3 ds
   \geq\gamma J(t,r).
\ee
We insert the estimates (\ref{Sest}), (\ref{kintest}), and (\ref{wrongjest})
into (\ref{dds3}) and use the fact that $r\geq R_-(t) \geq r_-$
along the characteristic under consideration to obtain the estimate
\begin{eqnarray*}
   \frac{d}{dt}\left[\left(1-\frac{2 M}{r}\right) E^2\right]
   & \ge & \frac{2\alpha}{r^2} J
   \biggl[2 \gamma w^2- \left(1+\frac{L_+}{W_-}\frac{1}{r_-^2}\right) E w
   -\frac{9}{2}\frac{a_+^2  M}{r_-}\,E w
   \\ & & \hspace{12em} -\left(3 M w^2 +\frac{L_+ }{r_-}\right)
   \frac{a_+ }{r_-}\biggr].
\end{eqnarray*}
Due to the estimate $E w \leq E^2 = 1+w^2 +L/r^2$ this implies that
\begin{eqnarray*}
\frac{d}{dt}\left[\left(1-\frac{2 M}{r}\right) E^2\right]
&\geq&   
\frac{2\alpha}{r^2} J\biggl[\left(2\gamma-1
   -\frac{9}{2}\frac{a_+^2  M}{r_-}-\frac{L_+ }{W_- r_-^2}
   -\frac{3a_+  M}{r_-}\right) w^2
\\ 
&&
{} -\bigg(1+\frac{L_+ }{W_- r_-^2}+\frac{9}{2}\,\frac{a_+^2  M}{r_-}\bigg)
   \bigg(1+\frac{L_+ }{r_-^2}\bigg)-\frac{L_+ a_+ }{r_-^2}\biggr].
\end{eqnarray*}
If we recall the definition of $\gamma$ in (\ref{gamma-def}) 
we see that the condition
(\ref{master1}) implies that the factor multiplying
$w^2$ in the right hand side of this estimate is positive.
Hence the estimate is still true if we replace $w$ by $W_-$,
since by assumption $w^2>W_-^2$ on the time interval $[0,t_0[$.
Again recalling the condition (\ref{master1}) we conclude that
\[ 
\frac{d}{dt}\left[\left(1-\frac{2 M}{r}\right) E^2\right] > 0
\]
as long as $t\in[0,t_0[$.
Hence for any characteristic in $\supp f$ on this time interval,
\[
\left(1-\frac{2 M}{r(t)}\right)\left(1+w^2(t)+\frac{L}{r^2(t)}\right)
\ge
\left(1-\frac{2 M}{r(0)}\right)\left(1+w^2(0) + \frac{L}{r^2(0)}\right).
\]
Thus
\[ 
w^2(t)+\frac{L}{r^2(t)}
   \geq\left(1-\frac{2 M}{r(0)}\right)\left(w^2(0)+\frac{L}{r^2(0)}\right)
   -\frac{2 M}{r(0)}\,. 
\]
Since $\dot{r}\ge 0$, $r(t)\ge r(0) > r_-$. Together with $w(0)\ge w_-$
this implies that
\begin{eqnarray*}
   w^2(t) & \ge & \left(1-\frac{2M}{r_-}\right) w_-^2
   -\frac{2 M}{r_-}\left(1+\frac{L_+}{r_-^2}\right)
   \\ & = & (1-\xi)\, w_-^2
   -\xi\,\left(1+\frac{L_+}{r_-^2}\right)>W_-^2\,,
\end{eqnarray*}
the latter because of (\ref{master2}). Therefore $w(t)>W_-$ for $t\in [0, t_0]$.
From (\ref{betaeq}) it follows that
\[ 
\beta(t,r) \geq - \frac{a_+}{r} \int_0^r 4\pi \jmath s^2\,ds
   \ge -\frac{a_+  M}{r_-}. 
\]
Thus by (\ref{charsys1}),
\[ 
\frac{\alpha_-}{a_+} \frac{W_-}{\sqrt{1+W_-^2 +L_+/r_-^2}}
   \leq \dot r \leq 1+\frac{a_+  M}{r_-}. 
\]
Since $r(0)\in ]r_-, r_+[$, this implies that $R_-(t)<r(t)<R_+(t)$
for $t\in [0, t_0]$; recall (\ref{R-def}) and (\ref{R+def}).
Therefore (\ref{rwbd}) holds on $[0, t_0]$,
and $t_0=T$. 

\smallskip

\noindent
{\em Step 2: Global existence in maximal-areal coordinates}.\\
The next step in the proof of the theorem is to show
that the estimates obtained in Step~1 on the 
maximal existence interval $[0,T[$ imply that $T=\infty$.

In \cite[Thm.~2.1]{Rl7} sufficient conditions 
for global existence in maximal-isotropic coordinates are obtained.
It is straightforward to see that these conditions
are also sufficient in our case, i.e., if we can show that
the support of the momenta and the metric function $a$
do not blow up, then $T=\infty$ follows; the 
local well-posedness in maximal-areal coordinates and continuation of
solutions
will be studied more systematically in \cite{AKR}.
At points $(t, r, \theta, \phi)$ with $r<R_{-}(t)$
there is no matter. Hence by (\ref{kap-form}), (\ref{alphar1}),  (\ref{ee1}),
and (\ref{bdryc}) 
this implies that 
\begin{equation} \label{vacuum}
a_r(t,r) = \alpha_r (t,r)=\kappa(t,r) = 0\ \mbox{for}\ r<R_-(t).
\end{equation}
In particular the bounds in (\ref{gote}) hold
for all $r\geq 0$ and $t \in [0, T[$. 
In view of the latter
it is sufficient to verify that 
\begin{equation} \label{Ebound}
\sup\{ E \mid (r, w, L)\in \supp f(t),\ t\in [0,T[ \} < \infty.
\end{equation}
In what follows $C>0$ denotes a constant which only depends
on the parameters specified in (\ref{master1}) and (\ref{master2})
and which may change from line to line.
By (\ref{E}) and (\ref{charsys1})--(\ref{charsys3}) it follows that
along characteristics,
\begin{equation} \label{dotE}
\dot E=-\frac{\alpha_r w}{a}
   -\frac{\alpha\kappa}{E}\Big(2w^2-\frac{L}{r^2}\Big). 
\end{equation}
Since by (\ref{rwbd}) and (\ref{R-def}) $r(t)\geq R_-(t)\geq C(1+t)$ for any 
characteristic in $\supp f$,
(\ref{kappa-est}) and (\ref{alpha-r}) imply that
\[
   |\kappa(t, r)|+
   \alpha_{r}(t, r) \leq  \frac{C}{(1+t)^2}.
\]
We point out that in this particular case when $w>0$ the first term of 
the right hand side of (\ref{dotE}) can be dropped since it is non-positive 
but this more general estimate will be needed later. 
Since $\alpha\le 1$ and $a\geq 1$ we thus obtain the estimate
\begin{equation} \label{Edotest}
|\dot E|\leq C \frac{E}{(1+t)^2}, 
\end{equation}
and Gronwall's lemma implies the desired bound  (\ref{Ebound}).
Thus $T=\infty$.

\smallskip

\noindent
{\em Step 3: Future geodesic completeness.}\\
In order to analyze this question polar coordinates such
as we use on the spatial $t=\mathrm{const}$ surfaces are not convenient,
and we introduce new coordinates
\[
x^0 = t,\ x^1 = r\sin\theta\cos\phi,\ x^2 = r\sin\theta\sin\phi,\
x^3 = r \cos\theta,
\]
which can be thought of as the corresponding Cartesian coordinates.
In these coordinates the metric becomes
\[
g_{00} = - \alpha^2 + a^2 \beta^2,\ g_{0i} = a^2 \beta
\frac{x_i}{r},\
g_{ij} = \delta_{ij} + (a^2-1)\frac{x_i x_j}{r^2},
\]
where Latin indices $i,j$ run from $1$ to $3$ and $x_i = \delta_{ij}x^j$.
Let us now consider 
an arbitrary future directed, time-like or null geodesic,
i.e., a solution $(x^\gamma (s),p^\gamma(s))$ of the geodesic equations
\begin{eqnarray*}
   \frac{dx^{\gamma}}{ds} & = & p^{\gamma}, \\
   \frac{dp^{\gamma}}{ds} & = & - \Gamma^{\gamma}_{\delta\epsilon}\,p^{\delta}p^\epsilon,
\end{eqnarray*}
where Greek indices $\gamma,\delta,\epsilon$ run from $0$ to $3$,
$\Gamma^{\gamma}_{\delta\epsilon}$ are the Christoffel symbols,
and 
\[
p^0 >0,\ g_{\gamma\delta} p^\gamma p^\delta = - m^2 \leq 0;
\]
$m^2$ is conserved along any geodesic, and $p^0$ cannot change sign.
Such a geodesic exists on a maximally 
extended interval $[0,s_+[$,
and future geodesic completeness means that $s_+=\infty$
for all such geodesics.

The following relations between the variables $r$, $w$, $L$, and $p^\gamma$ hold:
\begin{eqnarray*}
E
&=& \alpha \, p^0,\\
w
&=&
\left(\frac{x_i p^i}{r} + \beta p^0 \right)\, a,\\
\frac{L}{r^2}
&=&
\delta_{ij} p^i p^j - \left(\frac{x_i p^i}{r}\right)^2,
\end{eqnarray*}
where we now re-define
\[
   E=E(r,w,L)=\sqrt{m^2+w^{2}+L/r^{2}}.
\]
That we had $m=1$ before means that our system describes
an ensemble of particles of rest mass $1$, but for geodesic completeness 
we have to consider any time-like or null geodesic, i.e., we have to 
allow any $m\geq 0$.
Since $dt/ds = p^0 > 0$ we can re-parameterize the geodesic
by coordinate time $t\in [0,t_+[$. Along the geodesic
again (\ref{dotE}) holds.
By (\ref{vacuum}), $\dot{E}=0$ as long as $r \leq R_-(t)$ so that 
also (\ref{Edotest}) holds along the geodesic, and
$E\le C$ for some constant $C$.
Since by (\ref{gote}) $\alpha$ is bounded from below,
$p^0$ remains bounded on $[0,t_+[$. Since both $a$ and $\beta$ are
bounded as well also the $p^i$ remain bounded, and hence $t_+=\infty$. 
But since $dt/ds = p^0$ is bounded, this implies that
$s_+=\infty$, and
the proof of the theorem is complete. {\hfill$\Box$}

\end{document}